\documentclass[prd,aps,floats,twocolumn,showpacs]{revtex4}
\usepackage[dvips]{epsfig}

\begin{document}

\def\sqr#1#2{{\vcenter{\hrule height.#2pt
   \hbox{\vrule width.#2pt height#1pt \kern#1pt
      \vrule width.#2pt}
   \hrule height.#2pt}}}
\def\square{{\mathchoice\sqr64\sqr64\sqr{3.0}3\sqr{3.0}3}}

\title{Lagrangian with U(1)~--~SU(2) mixing}

\author{Bernd A.\ Berg}

\affiliation{Department of Physics, Florida State
             University, Tallahassee, FL 32306-4350, USA}

\date{November 18, 2009.} 

\begin{abstract}
Principal axis transformation is performed for a Lagrangian with
a  U(1)~--~SU(2) mixing term, that can cause a SU(2) deconfining 
transition.
\end{abstract}
\maketitle

Recently, it was shown in a lattice gauge theory simulation \cite{Be09}
that a U(1)~--~SU(2) interaction term can cause a SU(2) deconfining 
phase transition quite similar to the confinement-Higgs transition 
observed in \cite{Early}. This interaction requires unusual gauge 
transformations, which are written down in \cite{Be09a} for the 
continuum formulation. In the present note the results of a
principal axis transformation of this Lagrangian are given.

In the following we use Euclidean notation. The SU(2)$\otimes$U(1) 
Lagrangian of \cite{Be09,Be09a} reads
\begin{eqnarray} \label{L}
  L &=& -\frac{1}{2}{\rm Tr}\left(F^a_{\mu\nu}F^a_{\mu\nu}\right)
        -\frac{1}{2}{\rm Tr}\left(F^b_{\mu\nu}F^b_{\mu\nu}\right)
        \\ \nonumber &~& - \lambda\,{\rm Tr}
        \left(F^{\rm int}_{\mu\nu}F^{\rm int}_{\mu\nu}\right)
  \\ \label{Fa} F^a_{\mu\nu} &=& \partial_{\mu}A_{\nu} 
                              -  \partial_{\nu}A_{\mu}\,,      
  \\ \label{Fb} F^b_{\mu\nu} &=& \partial_{\mu}B_{\nu} -
  \partial_{\nu}B_{\mu} + ig_b\left[B_{\mu},B_{\nu}\right]\,, 
  \\ \label{Fint} F^{\rm int}_{\mu\nu} &=& g_a\,\partial_{\mu}A_{\nu} 
                                      - g_b\,\partial_{\nu}B_{\mu}\,.
\end{eqnarray}
where $A_{\mu}$ are U(1) fields written as $2\times 2$ matrices and 
$B_{\mu}=\vec{\tau}\cdot\vec{b}_{\mu}/2$ with $\tau_i$, $i=1,2,3$ the 
Pauli matrices are SU(2) fields. The ``diagonal gauge'' \cite{Be09a} 
is used in which the $A_{\mu}$ fields are given by $A_{\mu}=\tau_0
a_{\mu}/2$ with $\tau_0$ the $2\times 2$ unit matrix), so that
\begin{equation} \label{dgauge} 
  \left[A_{\mu},A_{\nu}\right] = \left[A_{\mu},B_{\nu}\right] = 0
\end{equation}
holds in this gauge. \smallskip

With the definitions ($i=1,2,3)$
\begin{eqnarray} 
  \label{Y1} Y^1_{\mu} &=& a\,A_{\mu}+b\,B_{\mu}\,,\\ 
  \label{Y2} Y^2_{\mu} &=& b\,A_{\mu}-a\,B_{\mu}\,,\\ 
  \label{Y3} Y^3_{\mu} &=& a\,A_{\mu}-b\,B_{\mu}\,,\\ 
  \label{Fi} F^i_{\mu\nu} &=& 
             \partial_{\mu}Y^i_{\nu}-\partial_{\nu}Y^i_{\mu}
             + ic_i\left[Y^i_{\mu},Y^i_{\nu}\right]\,.
\end{eqnarray}
the Lagrangian (\ref{L}) can be written in the principal axis form 
\begin{eqnarray} \label{Lp} 
  L &=& -\sum_{i=1}^3 \lambda_i\,{\rm Tr}
         \left(F^i_{\mu\nu}F^i_{\mu\nu}\right)~~~~{\rm with} 
  \\ \label{lambda} \lambda_1 &=&1+\lambda_3,~~\lambda_2~=~1\,,
     ~~\lambda_3~=~\frac{\lambda}{2}\left(g_a^2+g_b^2\right)\,,
  \\ \label{ci} c_1 &=& \frac{g_b}{\lambda_1 b}\,,~~c_2~=
     -\frac{g_b}{a}\,,~~c_3~=~\frac{g_b}{\sqrt{\lambda_1}\,b}\,,
  \\ \label{ab} a &=& \frac{g_a}{\sqrt{2(g_a^2+g_b^2)}}\,,~~
                b ~=~ \frac{g_b}{\sqrt{2(g_a^2+g_b^2)}}\,.
\end{eqnarray}
The algebra is verified by the FORM \cite{Form} program given in the 
appendix. The $Y^3_{\mu}$ fields are not independent, but can 
be expressed in terms of $Y^1_{\mu}$ and $Y^2_{\mu}$ as
\begin{eqnarray} \label{Y3Y1Y2}
  Y^3_{\mu} = \frac{a^2-b^2}{a^2+b^2}\,Y^1_{\mu} 
            + \frac{2ab}{a^2+b^2}\,Y^2_{\mu}\,,
\end{eqnarray}
so that $F^3_{\mu\nu}$ facilitates an interaction between $Y^1_{\,u}$ 
and $Y^2_{\,u}$.

It is tempting to identify $Y^1_{\mu}$ with the photon field 
$A^{\gamma}_{\mu}$, and $Y^2_{\mu}$ with the intermediate $Z$ boson 
field $Z_{\mu}$, where $\tau_0$ would become the hypercharge operator 
$Y$~\cite{Qu83}. However, it is not obvious that this can be done. 
Information about the true vacuum state is needed, which is in the 
standard model given by the expectation value of the initial Higgs 
field, so that mass terms for the $Z$ and $W$ bosons become explicit. 
As the deconfinement mechanism discovered in \cite{Be09} is 
non-perturbative, the Lagrangians (\ref{L}) and (\ref{Lp}) show 
no signs of this transition, which is encountered when $\lambda$ 
in (\ref{L}) is varied. Therefore, one has to rely on non-perturbative 
mass spectrum calculations and a detailed comparison of the spectral 
properties of our model with those of the electroweak Higgs model on 
the lattice \cite{Early} promises insights. Certainly it remains 
remarkable that the simple mixing term (\ref{Fint}) has the ability 
to drive the SU(2) part of the theory from the confined into the 
deconfined phase.

\acknowledgments 
This research was in part supported by the DOE grant DE-FG02-97ER41022 
and by a Humboldt Research Award. Some of the work was done at Leipzig 
University and I am indebted to Wolfhard Janke and his group for their 
kind hospitality. 

\appendix \section{Transformation of the Lagrangian.}

The algebra, which leads from (\ref{L}) to (\ref{Lp}) is
verified by the FORM \cite{Form} program listed in the 
following.

\begin{small} \begin{verbatim}
* Bernd Berg Nov 13 2009. 
* Transformation of the SU2xU1 Lagrangian.
* sqrtx=sqrt(2*(ga^2+gb^2)), CB=[Bu,Bv].
     Symbol Fa,duAv,dvAu, ga,i,gb,la;
     Symbol a,b,sqrtx, l3,l2,l1 c1,c2,c3;
     Function Fb,Fint,duBv,dvBu, CB;
     Function F1,F2,F3,duY1v,dvY1u,duY2v,dvY2u,duY3v,dvY3u;
     Off statistics;
     Local L=-Fa*Fa/2-Fb*Fb/2-la*Fint*Fint;
     Local Ld=-l1*F1*F1-l2*F2*F2-l3*F3*F3;
     Local ZLL=sqrtx^2*l1^2*(L-Ld);
     Local ZY3=(a^2+b^2)*dvY3u-(a^2-b^2)*dvY1u-2*a*b*dvY2u;
     id Fa=duAv-dvAu;
     id Fb=duBv-dvBu+i*gb*CB;
     id Fint=ga*duAv-gb*dvBu;
     id F1=duY1v-dvY1u+i*c1*b^2*CB;
     id F2=duY2v-dvY2u+i*c2*a^2*CB;
     id F3=duY3v+dvY3u+i*c3*b^2*CB;
     id duY1v=+a*duAv+b*duBv;
     id dvY1u=+a*dvAu+b*dvBu;
     id duY2v=+b*duAv-a*duBv;
     id dvY2u=+b*dvAu-a*dvBu;
     id duY3v=+a*duAv-b*duBv;
     id dvY3u=+a*dvAu-b*dvBu;
     id CB*duAv=duAv*CB;
     id CB*dvAu=dvAu*CB;
     id dvAu*CB=-duAv*CB;
     id dvBu*CB=-duBv*CB;
     id CB*dvBu=-CB*duBv;
     id c1=+gb/b/l1;
     id c2=-gb/a;
     id c3=+gb*sqrt_(1/l1)/b;
     id sqrt_(1/l1)^2=1/l1;
     id a=ga/sqrtx;
     id b=gb/sqrtx;
     id sqrtx^2=2*(ga^2+gb^2);
     id l1=1+l3;
     id l2=1;
     id l3=la*(ga^2+gb^2)/2;
     id dvAu^2=duAv^2;
     id dvBu*dvBu=duBv*duBv;
     id dvBu*duAv=duBv*dvAu;
     Print;
     .end
\end{verbatim} \end{small} \smallskip

With the substitutions corresponding to identities given in the main 
text, the program calculates $ZLL=0$ and $ZY3=0$. Respectively, this 
proves the equality of Eq.~(\ref{L}) and (\ref{Lp}), and the validity 
of Eq.~(\ref{Y3Y1Y2}) for $Y^3_{\mu}$. Symbols are commuting and
functions are non-commuting objects in FORM.

\end{document}